\documentstyle[preprint,eqsecnum,aps]{revtex}

\begin{document}
\draft
\preprint{NHCU--HEP--95/1-T}
\title{\large
Exact Solution of a One Dimensional Deterministic Sandpile Model
}\vskip 1cm
\author{Darwin Chang$^{(1,2)}$
Shih-Chang Lee$^{(2)}$, and
Wen-Jer Tzeng$^{(3)}$}
\address{$^{(1)}$Department of Physics,
National Tsing-Hua University, Hsinchu, Taiwan  30043, R.O.C.\\
$^{(2)}$Institute of Physics, Academia Sinica, Taipei, Taiwan 115, R.O.C.\\
$^{(3)}$Department of Physics,
National Chung-Hsin University, Taichung, Taiwan, R.O.C.}
\maketitle
\begin{abstract}
Using the transfer matrix method,
we give the exact solution of a deterministic sandpile model
for arbitrary $N$, where $N$ is the size of a single toppling.
The one- and two-point functions are given in term of the eigenvalues of
an $N \times N$ transfer matrix.  All the n-point functions can be found in
the same way.   Application of this method to a more
general class of models is discussed.  We also present a quantitative
description of the limit cycle (attractor) as a multifractal.
\end{abstract}
%
%
\narrowtext
\section{Introduction}
Sandpile model has been conceived as the simplest model which
can illlustrate self-organized criticality(SOC).  It has developed into an
interesting cellular automata model of nonlinear dynamics in its own right.
Bak, Tang, and Wiensenfeld\cite{BTW,bak88} are the first to use numerical
method to study
some of the sandpile models and observed that the models automatically
evolve into a self-organized critical state while they possess $1/f$
spectra in both spatial and temporal distributions of certain physical
quantities.  It was conjectured that SOC maybe an universal characteristic
underlying the nonlinear, dispersive systems in nature such as earthquakes,
forest fire, turbulence, etc...

The simplest cellular automata model of sandpile is to assign
a height number $h_i$ to each site on a
one dimensional lattice with length $L$.
There are two basic operations of the model \-- dropping and toppling.
Dropping means that one sand is added at some site of the lattice,
i.e. $D_i: h_i\rightarrow h_i+1$.  Toppling occurs when a slope
(defined as the difference in height between adjacent sites) exceeds
some critical value, $N$. If toppling occurs at one site, some sands at the
site will be moved to the other sites which may trigger further
topplings.

Even with these simple rules of evolution, the analytic solution of
the sandpile model are typically very difficult to come by especially when
$L$ or $N$ becomes large, or when the lattice dimension is more than one.
Most researcher handle the models by
numerical simulation\cite{NS1,NS2,NS3}.  If the rules are
such that the evolution of the system is independent of the order of
the droppings, then the model is called abelian.  For a large class of
abelian models some exact results have been obtained by Dhar et
al.\cite{abelian}  The non-abelian models are typically much harder to
solve.  Here we shall investigate a class of nonabelian models in the
deterministic limit, i.e., the sand being dropped at a fixed
site.

We consider the one
dimensional case and label the sites from left to right as 1 to $L$.
The sand is dropped only at the site 1. If the slope at a site exceeds
a given number $N$, then the sand will topple to the right. Let the slope
$\sigma_i=h_i-h_{i+1}$, then the toppling rule is ``if $\sigma_i>N$,
then $\sigma_{i-1}\rightarrow\sigma_{i-1}+N$,
$\sigma_i\rightarrow\sigma_i-(N+1)$ , and
$\sigma_{i+N}\rightarrow\sigma_{i+N}+1$''. The rule should be modified
when toppling occurs near the boundary.  The condition at the left
boundary is trivial.  When sands reach beyond the right boundary they drop
out from the system, i.e. we keep $h_i=0$ for $i>L$. A state in which
all $\sigma_i\leq N$ is called a stable state.  Toppling stops when a
stable state is reached.  Each dropping and subsequent toppling
processes will result in the transition from one stable state to
another.  Since there are only a
finite number of different states in the system, after dropping enough
sand at site 1, the system will step into a cycle called the limit
cycle.  For the system that we consider here there is only one limit
cycle in the problem.  Not all the stable states are in the limit cycle.
Those in the limit cycle are called allowed states.
In the following, all the states we will refer to
are allowed states.
The number of allowed states in the limit
cycle is $N^L$.\cite{Lee1}

Some special solvable non-abelian models have been investigated in
Ref.\cite{Lee1,Lee2,ccl}.   In Ref.\cite{Lee1}, the simplest ($N=2$) of
a class of non-abelian sandpile models was solved.
Some general property of the model for arbitrary $N$ was also
exposed including some results about the random drop cases\cite{Lee2}.
In Ref.\cite{ccl} a new method using the transfer matrix idea from the
statistical mechanics models was developed and applied to solve the $N=3$
case of the same class of sandpile models.  It was commented that the
method should also be useful for solving the models with arbitrary $N$.
In this note, we wish to develop this idea further and supply a
solution of the model with arbitrary $N$.

The method to be introduced here is particularly
useful for those models in which the
structure of the limit cycles has been worked out.  Once the cycle structure
is known, its information can be succinctly summarized in a matrix
which we call the transfer matrix by analogy with the similar matrix in
statistical mechanics.  For the deterministic model defined above, the
structure of the limit cycle has been worked out in ref.\cite{Lee1}.
Therefore we should use it as our main example, the method may be
applicable to a much wider classes of models.  A multifractal description of
the limit cycle is also worked out.

\section{The definition of Transfer matrix}
A allowed states of length $L$ in the limit cycle are characterized by three
conditions: \begin{enumerate}
\item $\sigma_i \ge 0$ for $0<i\le L$.\label{enum:constrain1}
\item There exists at least one site $i$ for any consecutive $N$ sites such
that $\sigma_i=N$.(stability condition)\label{enum:constrain2}
\item There exists a site $i$ satisfying $L-\sigma_L\leq i<L$ such that
$\sigma_i=N$.(boundary condition)\label{enum:constrain3}
\end{enumerate}

If $[\sigma_1 \sigma_2...\sigma_L]$ is an allowed state in the limit cycle
with length $L$, then $[\sigma_{i+1} \sigma_{i+2}...\sigma_L]$ is also an
allowed state in the limit cycle with length $L-i$.

\subsection{Reduced basis
}

To make the method of tranfer matrix more powerful, it is useful to
define a reduced basis,
$||n\rangle_l$, which classifies the allowed states according
to the constraint on their preceding sites (to their left).
The set of allowed states of length $l$ can be classified according to a
quantum number $n$, $1\leq n\leq N$.
A state $[\eta_1\eta_2...\eta_l]$,
which is an allowed state in the limit
cycle with length $l$, is said to have the quantum number $n$ if
$\eta_{N+1-n}=N$ and $\eta_i<N$ for $i<N+1-n$ when
$N+1-n\leq l$, while $\eta_l=l+n-1$ and $\eta_i<N$ for $i<l$ when $N+1-n>l$.
A collection of such states can be denoted as $||n\rangle_l$
and called a reduced state, while $|n\rangle_l$ denotes a typical state in
$||n\rangle_l$.
Every allowed state belongs to one of the reduced states.
We therefore call the set of $||n\rangle_l$, with $1\leq n\leq N$, a
reduced basis.
The quantum number $n$ labels the fact that, for any state
$[\eta_1\eta_2...\eta_l]$ in $||n\rangle_l$, one of the $n$ slopes to the
left of $\eta_1$ has to be $N$.

For a allowed state of length $L$, $[\sigma_1 \sigma_2...\sigma_L]$, the
last $l$ slopes $[\sigma_{L-l+1} \sigma_{L-l+2}...\sigma_L]$ is an allowed
state of length $l$, which belongs to some reduced basis $|n\rangle_l$, and
$[\sigma_1\sigma_2...\sigma_{L-l}\sigma'_{L-l+1}(=n)]$ is an allowed state
of length $L-l+1$.  This correspondence is one-to-one for fixed $l$ and this
provides a route to construct the limit cycle of length $L$ from the limit
cycle of length $l(<L)$.

\subsection{
The time-average of space dependent functions
}
Let $S_l$ be the number of allowed states of length $l+1$ with $\sigma_{l+1}
=N$.  From the three conditions for allowed states in the limit cycle, we have
$S_l=(N+1)^l$ for $0 \leq l <N$, $S_N=(N+1)^N-N^N$, and $S_{l+1}=
(N+1)S_l-N^NS_{l-N}$ for $l>N$.
For any allowed state of length $L$ with
$\sigma_i=N$, $[\sigma_{i+1} \sigma_{i+2}...\sigma_L]$ can be any allowed
state of length $L-i$.  The number of allowed states of length $L$ with
$\sigma_i=N$ is thus $N^{L-i}S_{i-1}$.

It should be noted that the three conditions for allowed states in the limit
cycle only differentiate $\sigma_i=N$ and $\sigma_i\neq N$ unless $i=L$.
Since we know the total degeneracy of states for $\sigma_i$ taking
arbitray value and the degeneracy for states in which $\sigma_i = N$ can
be calculated using $S_l$, the degeneracy for $\sigma_i \neq N$ can be
easily obtained.  Therefore, $S_l$ contains all the information about the
time average of
space dependent functions, such as the n-point function
$<\sigma_i...\sigma_j>$.  Here $<...>$ means the average
in time over a limit cycle.  The one-point function can be evaluated
in terms of the summation of all allowed states in a limit cycle, i.e.,
\begin{eqnarray}
<\sigma_i>&&={NN^{L-i}S_{i-1}+{N-1\over 2}(N^L-N^{L-i}S_{i-1})\over N^L}
\nonumber \\
&&={N-1\over 2}+{N+1\over 2}{S_{i-1}\over N^i},~~~~~~~~~~i<L.\label{eqn:1pf}
\end{eqnarray}
Similarly, the two-point function is
\begin{eqnarray}
<\sigma_i\sigma_j>=&&{(N-1)^2\over 4}+{N^2-1\over 4}\left({S_{i-1}\over N^i}
+ {S_{j-1}\over N^j}\right)\nonumber \\
&&+{(N+1)^2\over 4}{S_{i-1}S_{j-i-1}\over N^j},~~~~~~~~~~i<j<L.\label{eqn:2pf}
\end{eqnarray}
Actually, all the n-point functions can be expressed in a similar form.
Therefore, the calculations of the n-point functions are reduced to the
calculation of $S_l$.
It should be noted that these time-average of space dependent functions are
independent of $L$.

One can use mathematical induction to get
the following close form for $S_l$.
\begin{eqnarray}
S_{mN+l}=&&(N+1)^{mN+l}\nonumber\\
&&+\sum_{j=1}^m(-1)^j(N+1)^{(m-j)N-j+l}N^{jN}\left[
{(m-j)N+l\choose j}+(N+1){(m-j)N+l\choose j-1}\right] \label{eqn:sl}
\end{eqnarray}
with $m\geq 0$ and $0\leq l<N$.
However, such tedious close form is difficult to use.   A better
form can be obtained by transfer matrix method.

\subsection{Transfer matrix
}

For any state in $|n\rangle_l$, we can add a preceding slope $\sigma=N$,
resulting in a state in $|N\rangle_{l+1}$, or add a preceding slope
$\sigma=0,1,2,...,N-1$ if $n>1$, resulting in a state in $|n-1\rangle_{l+1}$.
We can express this property in term of
a transfer operator $T$ such that $T||n\rangle_l =
||N\rangle_{l+1} +N||n-1\rangle_{l+1}$, if $n>1$, and $T||1\rangle_l =
||N\rangle_{l+1}$.
For example, the first equation represents the fact that, given a
typical state, $[\eta_1\eta_2...\eta_l]$,
in $||n\rangle_l$ the corresponding allowed state
$[\eta,\eta_1\eta_2...\eta_l]$ contains one state in $||N\rangle_{l+1}$
and $N$ states in $||n-1\rangle_{l+1}$.
As far as transfer matrix is concerned, all the allowed states within the same
reduced state are identical and there is no need to distinguish them.
It is also clear from the definition of transfer matrix that its
properties are independent of the subscript $l$.  Since it is obvious that
each operation of $T$ increases the $l$ by one, the subscript can be
ignored all together.

Represent $||N\rangle$ by $col(1~0~...~0)$,
$||N-1\rangle$ by $col(0~1~0~...~0)$,
..., $||1\rangle$ by $col(0~...~0~1)$, the transfer operator can thus be
represented as an $N \times N$ matrix:
\begin{equation}
T=\pmatrix{1&1&1&\ldots&1&1\cr
           N&0&0&\ldots&0&0\cr
           0&N&0&\ldots&0&0\cr
           \vdots&\vdots&\vdots&\ddots&\vdots&\vdots\cr
           0&0&0&\ldots&0&0\cr
           0&0&0&\ldots&N&0\cr}.
\end{equation}
Note that $col(1~1~...~1)$ is an eigenvector of $T$ of eigenvalue $N$.
The allowed states of length one are $\sigma_1=1,2,...,N$, which correspond
to the reduced basis $||1\rangle_1$, $||2\rangle_1$, ..., $||N\rangle_1$,
respectively.  The number of allowed states of length $L$ is easily proved
to be $(1~1~...~1)T^{L-1}col(1~1~...~1)=N^L$.  This is the simplest way of
deriving this result pointed out already in ref.\cite{Lee1}.
The number of states corresponds to the reduced basis $|n\rangle_l$ with
$1\leq n\leq N$ is
$(\delta_{N,n}~\delta_{N-1,n}~...~\delta_{1,n})T^{l-1}col(1~1~...~1)
=N^{l-1}$, which is independent of $n$,
and
\begin{equation}
S_l=(1~1~...~1)T^{l}col(1~0~...~0).
\end{equation}

$S_l$ can now be expressed in terms of the eigenvalues of the transfer matrix
$T$, and we have
\begin{equation}
S_l={N^{l+1}\over N+1}\left(2+\sum_{m=1}^{N-1}\lambda_m^{i+1}\right)
\end{equation}
where $N\lambda_0(\lambda_0=1)$, $N\lambda_1$, ..., $N\lambda_{N-1}$ are the
eigenvalues of $T$ and $\lambda_1$, $\lambda_2$, ..., $\lambda_{N-1}$ are
the solutions of the equation
\begin{equation}
1+2\lambda+3\lambda^2+\cdots+N\lambda^{N-1}=0.\label{lambda}
\end{equation}
The one and two-point functions are thus
\begin{equation}
<\sigma_i>={N+1\over
2}-{1\over2}\sum_{m=1}^{N-1}\lambda_m^i,~~~~~~~~~~i<L\label{onept}
\end{equation}
and
\begin{eqnarray}
<\sigma_i\sigma_j>=&&{(N+1)^2\over 4}+{N+1\over 4}\sum_{m=1}^{N-1}
\lambda_m^i+{N-1\over 4}\sum_{m=1}^{N-1}\lambda_m^j\nonumber \\
&&+{1\over 2}\sum_{m=1}^{N-1}\lambda_m^{j-i}+{1\over 4}
\sum_{m=1}^{N-1}\lambda_m^{i}
\sum_{k=1}^{N-1}\lambda_k^{j-i},~~~~~~~~~~i<j<L.\label{twopt}
\end{eqnarray}
For $N=2$ and $3$, the results agree with \cite{Lee1,ccl}.

With $y=1$/$\lambda$ and multiplying Eq.~(\ref{lambda}) by $y-1$, we have
\begin{equation}
y^N+y^{N-1}+\cdots+y=N\Rightarrow |y|>1\Rightarrow |\lambda|<1
\end{equation}
where the redundant solution $y=1$ is excluded.  Multiplying by $y-1$ again,
we have
\begin{equation}
y^{N+1}=(N+1)y-N\Rightarrow (N+1)|y|+N\geq |y|^{N+1},\label{ymax}
\end{equation}
which implies $|y|\leq 2$
for $N\geq 2$ and therefore $|y|\leq (3N+2)^{1/(N+1)}$.  The solutions of
$y$ are inside the annulus $1<|y|\leq y_{max}=(3N+2)^{1/(N+1)}$.  In fact
for large $N$ one can show that $y_{max} \cong 1 + N^{-1} log (2N)$ and
therefore as $N\rightarrow \infty$ all the roots have $|y|$ near 1.  From
Eq.~(\ref{ymax}) we have
\[
|y|^{N+1}=(N+1)\left|y-{N\over N+1}\right|,
\]
which means the solutions with smaller $|y|$ are all near $y=1$ for large $N$.

To explore the asymptotic behavior of $S_l$, we are interested in the
solutions of Eq.~(\ref{lambda}) with larger $|\lambda|$.  Therefore, we
would like to find the solutions of Eq.~(\ref{ymax}) with smaller
$|y|$ excluding the redundant solution $y=1$.  Considering only one of the
complex conjugate solutions in the upper half of the complex plane, let $y_j
= a_je^{i\theta_j}$ with $0<j<N/2$ and $\phi_j=\angle[(N+1)y_j-N]$ with $0<
\phi_j<\pi$ such that $(N+1)\theta_j=\phi_j+2j\pi$.  It should be noted that
$(N+1)\theta_0=\phi_0$ results in the redundant solution $y_0=1$.  From
Eq.~(\ref{ymax}) we
have
\[
a_j^{2N+2}=(N+1)^2a_j^2+N^2-2N(N+1)a_j\cos\theta_j
\]
and
\[
\tan\phi_j={(N+1)a_j\sin\theta_j\over
(N+1)a_j\cos \theta_j-N}
\]
For large $N$ we only have to consider the case of $j/N\ll 1$ to find the
smaller $|y|$ solutions.  Let $a_j=1+x_j/(N+1)$ and $\theta_j= (2j\pi+
\phi_j)/(N+1)\ll 1$, we have (as $N \rightarrow \infty$)
\begin{equation}
e^{2x_j}-(1+x_j)^2 \cong (\phi_j+2j\pi)^2\label{xj}
\end{equation}
and
\begin{equation}
\tan\phi_j \cong {2j\pi+\phi_j\over 1+x_j}
\label{phij}
\end{equation}
where the trigonometric functions have been expanded to the second order.
Table \ref{xphi} shows the values of the first few $x$'s and $\phi$'s as $N
\rightarrow \infty$.  From Eq.~(\ref{twopt}), the correlation length is
found to be $N/x_1$ for large $N$.

\section{Multifractal Characterization of the limit cycle}
Now we would like to describe the limit cycle (attractor) quantitatively.
In general, the state in the limit cycle corresponds to a path in a tree.
It is now possible to identify a multifractal along the time axis.
Rescaling the time variable so that each time step is of length
$\delta=N^{-L}$, the duration for a limit cycle fits into the time interval
$[0,1]$.  Consider the states $[\sigma_1\sigma_2...\sigma_L]$ in
the limit cycle
which contains ${\bf N}_0$ $0$'s, ${\bf N}_1$ $1$'s, ..., and ${\bf N}_N$
$N$'s,.  The
corresponding time steps for such states form a fractal in the interval
$[0,1]$ with fractal dimension $f(\xi_0, \xi_1, ..., \xi_N)$  where
$\xi_i={\bf N}_i/L$, for $0\leq i\leq N$ and satisfy $\sum_i\xi_i=1$.  The
above
relations are exact as $L \rightarrow \infty$.

For $N=2$, it is straight forward to get
\begin{equation}
f(\xi_0, \xi_1, \xi_2)={1\over \ln2}\left[\xi_0\ln \left( {\xi_2-\xi_1-
\xi_0\over\xi_0}\right)+\xi_1\ln \left( {\xi_2-\xi_1-\xi_0\over\xi_1
}\right)+\xi_2\ln \left({\xi_2\over\xi_2-\xi_1-\xi_0}\right)\right].
\end{equation}

This function depends on $\xi_0$ and $\xi_1$ in the same way, which
suggests that the boundary effect due to the specific constraint on
$\sigma_L$ is negligible in the large $L$ limit.  Neglecting this boundary
effect, there is no more difference between $n=0$, $1$, ..., and $N-1$.
Thus it is natural to define the fractal dimension as a function of
$\xi={\bf N}_N/L$ with $1/N\leq\xi\leq 1$.  The problem is equivalent to
counting the ways to distribute $(1-\xi)L$ objects into $\xi L$ slots
and every slot admits at most $N-1$ objects.  Now we have
\begin{equation}
\left(N^L\right)^{f(\xi)}\cong {1\over(L-{\bf N}_N)!}{d^{L-{\bf N}_N}\over
dx^{L-{\bf N}_N}}
\left.\left({x^N-1\over x-1}\right)^{{\bf N}_N}\right|_{x=0}N^{L-{\bf
N}_N}\label{f}
\end{equation}
The ``$\cong$'' is ``$=$'' as $L\rightarrow\infty$.
After taking the differentiation with $x$ apart on $[(x^N-1)/(x-1)]^{{\bf
N}_N-1}$
and $1+x+x^2+...+x^{N-1}$, we have
\begin{equation}
N^{f+(1-\xi)f'}=\sum_{k=0}^{N-1}N^{k(1-f)+k\xi f'}\label{df}
\end{equation}
where $f(\xi+\triangle\xi)$ is approximated as $f(\xi)+\triangle\xi f'$.
Let $f(\xi)=g(\xi)/\ln N+1-\xi$
with $R\equiv e^{g'}$ and $Z\equiv e^{-(g-\xi g')}$, which means
\begin{equation}
{R'\over R}\xi={Z'\over Z}.\label{xi}
\end{equation}
Now Eq.~(\ref{df}) becomes
\begin{equation}
R=Z+Z^2+\cdots+Z^N.
\end{equation}
It follows from the definitions of $R$ and $Z$ that
\begin{equation}
g(\xi)=\xi\ln(Z+Z^2+\cdots+Z^N)-\ln Z\label{gz}
\end{equation}
with
\begin{equation}
\xi={1+Z+Z^2+\cdots+Z^{N-1}\over 1+2Z+3Z^2+\cdots+NZ^{N-1}}.\label{xiz}
\end{equation}
This result obviously satisfy the boundary conditions $f(\xi=1/N)=1-1/N$
and $f(\xi=1)=0$.  $f(\xi)=1$ and $f'(\xi)=0$ happen at $\xi=2/(N+1)$,
which is consistent with $<\sigma_i>\rightarrow (N+1)/2$ as $i \rightarrow
\infty$ from Eq.~(\ref{onept}).  Equations (\ref{gz}) and (\ref{xiz}) express
the fractal dimension $f(\xi)$ in terms of the variable $Z$, $0\leq Z \leq
\infty$.

\section{Discussions}
We  have introduced the concept of transfer matrix into the study of
steady properties  of sandpile  models. This  is closely related to the
Hamiltonian formulation of the usual statistical mechanics. Indeed, one
may regard the formulas for the correlation function such as equation
(\ref{eqn:1pf}) and (\ref{eqn:2pf}) as a ``path integral'' expressions
in a discrete formulation. The transfer matrix plays the role of the
evolution operator.

The usefulness of the transfer matrix formulation was illustrated by
deriving the one- and two-point correlation functions for a
deterministic sandpile model with an arbitrary critical slope $N$.
We found that the two-point function decreases exponentially
as the separation of the two point increases with  a correlation length
depending on $N$.  In the unit of lattice spacing, the correlation length is
found to be the largest (in magnitude) root of an ($N$-1)th order polynomial.
The correlation length is proportional to $N$ in the large $N$ limit.

In the $L \rightarrow \infty$ limit, the states in the limit cycle can be
characterized as a multifractal.  We derive an explicit expression for the
multifractal dimension for arbitrary $N$.

Even though there exists only short range spatial correlation for any $N$,
power law behavior is expected for the time-time correlation function.
Derivation of an explicit formula for this correlation function is under
investigation.

\acknowledgments

This work is supported in part by  grants from the National Science
Council  of Taiwan-Republic of China under the contract number
NSC-84-0208-M001-069, NSC-84-02112-M007-042, NSC-84-2112-M001-022,
NSC-84-02112-M007-016 and NSC-84-2112-M005-010.

\begin{table}
\caption{ Some numerical values of $x_j$ and $\phi_j$
as $N\rightarrow\infty$.\label{xphi}}
\begin{tabular}{rdd}
$j$&$x_j$&$\phi_j$\\
\tableline
1&2.089&1.178\\
2&2.664&1.312\\
3&3.026&1.374\\
4&3.292&1.410\\
5&3.501&1.435\\
\end{tabular}
\end{table}

\begin{references}
\bibitem{BTW} Per Bak, Chao Tang, and Kurt Wiesenfeld, Phys. Rev. Lett.
{\bf 59}, 381 (1987);
Chao Tang and Per Bak, Phys. Rev. Lett. {\bf 60} 2347 (1988).
\bibitem{bak88}P. Bak, C. Tang, and K. Wiesenfeld, Phys. Rev. {\bf A38}, 364
(1988).
\bibitem{NS1} Ashvin B. Chhabra, Mitchell J. Feigenbaum, Leo P, Kadanoff,
Amy J. Koolan Itamar Procaccia, Phys. Rev. E {\bf 47} 3099 (1993).
\bibitem{NS2} Leo P. Kadanoff, Sidney R. Nagel, Lei Wu, Su-min Zhou,
Phys. Rev. A {\bf 39} 6524 (1989).
\bibitem{NS3} Kurt Wiensenfeld, James Theiler, Bruce  McNamara,
Phys. Rev. Lett. {\bf 65} 949 (1990).
\bibitem{abelian} D. Dhar, R. Ramaswamy, Phys. Rev.  Lett. {\bf 63} , 1659
(1989); D. Dhar, Phys. Rev.  Lett. {\bf 64} 1613 (1990).
\bibitem{Lee1} S.-C. Lee, N. Y. Liang, and W.-J. Tzeng, Phys. Rev. Lett.
{\bf 67}, 1479 (1991); erratum-ibid. {\bf 68}:1442 (1992).
\bibitem{Lee2} S.-C. Lee and W.-J. Tzeng, Phys. Rev. A {\bf 45}, 1253
(1992).
\bibitem{ccl} Darwin Chang, C.-S. Chin and S.-C. Lee, Chin. J. Phys. {\bf
32}, 405 (1994).
\bibitem{RWA} J. Krug, J.E.S. Socolar, G. Grinstein,
Phys. Rev. A {\bf 46}, R4479 (1992).
\end{references}
\end{document}